\documentclass[12pt]{article}

\usepackage{amsmath}
\usepackage{graphicx}  
\usepackage{dcolumn}   
\usepackage{bm}        
\usepackage{amssymb}   
\usepackage{latexsym,amsfonts,amsmath,amsthm,epsfig,tabularx,hyperref}

\usepackage{comment}

    \usepackage{hyperref}

    \usepackage{algorithm,algorithmic}

\newtheorem{theorem}{Theorem}

\newcommand\e{\varepsilon}

\newcommand{\complex}{{\mathbb C}}

\newcommand{\norm}[1]{\| #1\|}
\newcommand{\ket}[1]{|#1\rangle}

\newcommand{\lr}\longrightarrow
\newcommand{\ra}\rightarrow

\title{A fast algorithm for approximating the ground state energy on a quantum computer}
\author{A. Papageorgiou, I. Petras, J. F. Traub and C. Zhang}
\date{\today }

\begin{document}
\maketitle
\begin{abstract} Estimating the ground state energy of a multiparticle system
with relative error $\e$ using deterministic classical algorithms has cost 
that grows exponentially with the number of particles. The problem
depends on a number of state variables $d$ that is proportional
to the number of particles and suffers from the curse of dimensionality.
Quantum computers can vanquish this curse. 
In particular, we study a ground state eigenvalue problem and 
exhibit a quantum algorithm that achieves
relative error $\e$ using a number of qubits
$C^\prime d\log \e^{-1}$ with total cost (number of queries plus
other quantum operations) $Cd\e^{-(3+\delta)}$,
where $\delta>0$ is arbitrarily small and $C$ and $C^\prime$ are independent
of $d$ and $\e$.

\end{abstract}

\section{Introduction}

A difficult and challenging problem in modern science is to
accurately compute properties of physical and chemical systems.
One of the difficulties in
carrying out precise calculations arises from the computational
demands the Schr\"odinger equation presents. The computational
resources needed to obtain accurate solutions appear to be exponential in the
size of the physical system. As a result these problems 
are considered intractable on a classical computer 
for systems that are not trivial in size. 
For an overview of the numerical methods used for the solution
of such problems
see \cite{LeBris,Lub} and the references therein.

So far there have been mixed results about the potential power of
quantum computers relative to that of classical computers. For some
problems, such as factoring large numbers, quantum computers offer
exponential speedups relative to the best classical algorithms known. On
the other hand, there are results about the limits of quantum
computation \cite{var}, as well as results showing that certain problems
are hard. For instance, estimating the ground state
eigenvalue of arbitrary local Hamiltonians is a QMA complete 
problem \cite{kit}. 

Although there are fundamental problems in complexity theory that remain open,
there is a distinct category of problems for which quantum computers can
offer substantial speedups relative to classical computers. This includes
problems, such as multivariate integration, path integration
and multivariate approximation, 
that suffer from the {\it curse of dimensionality} 
in the classical deterministic worst case. 
Quantum computers can vanquish the curse; 
see e.g. \cite{No01,NSW04, TW02}.  
R. E. Bellman introduced the term {\it curse of dimensionality}
referring to multivariate problems whose complexity grows exponentially
with the number of variables and so are impossible to solve
when the number of variables is large. 

An important problem in physics and chemistry that falls in this
category is the estimation of the ground state eigenvalue 
of a time-independent Hamiltonian 
corresponding to a multiparticle system.
Solving such problems on a classical computer
in the worst case has cost exponential in the number of
particles. In particular, the number of state variables $d$ is
proportional to the number of particles and the cost to
solve the problem with relative accuracy $\e$ may grow as $\e^{-d}$.
For these reasons researchers have been experimenting 
with quantum computers to solve eigenvalue problems in quantum chemistry 
with very encouraging results \cite{JDu10,nature}. See also 
\cite{Kassal08,Kassal10} and the references therein. 

We remark that recently there has been a fair amount of work 
dealing with eigenvalue problems
see, e.g. \cite{5,1,6,4,3,7,2}.
However, our results are different.
The other papers either address different 
eigenvalue problems, or use spin models, 
or study classical algorithms, or do not obtain
algorithm cost and error estimates.

In particular, we study a ground state eigenvalue problem and
we exhibit a quantum algorithm that achieves relative error $\e$ with cost
$Cd\e^{-(3+\delta)}$, where $\delta >0$ is an
arbitrarily small positive number.
The cost includes the number of queries plus
all other quantum operations.
The algorithm uses $C^\prime d\log \e^{-1}$ qubits.
The constants $C$ and $C^\prime$ as well as all
constants in our estimates throughout
this paper are independent of $d$ and $\e$.

We stress that we are not dealing with an arbitrary eigenvalue estimation 
problem. In our case we are able to obtain efficiently a rough but very useful 
approximation of the ground state eigenvector.
Abrams and Lloyd \cite{AL99} were the first to demonstrate the advantages 
of approximate eigenvectors in solving problems of physical
interest. 
Consequently, the cost to implement and simulate the evolution of the 
Hamiltonian for the amount of time prescribed by the accuracy demand
determines the cost 
to approximate the ground state eigenvalue.

We now consider the problem in more detail.
If the potential is a function of only state
variables then the ground state energy is given by the smallest eigenvalue
$E_1$ of the equation
\begin{eqnarray*}
(-\tfrac 12 \Delta +V)\Psi_1(x) &=& E_1\Psi_1(x)\quad \mbox{for all}\;\;
x\in I_d:=(0,1)^d, \\
 \Psi_1(x) &=& 0 \quad \mbox{for all}\;\; x\in  \partial I_d,
\end{eqnarray*}
where $\partial I_d$ denotes the boundary of the unit cube,
$x$ is the position variable, and $\Psi_1$
is a normalized eigenfunction. 
For simplicity we assume that all masses and the normalized Planck
constant are one. The boundary conditions are for particles in a box.
Multiparticle systems on bounded domains with the wave function equal to zero
on the boundary have been studied in the literature;
see e.g. \cite[p. 621]{LeBris}. 

This eigenvalue problem is called the time-independent Schr\"odinger
equation in the physics literature and the Sturm-Liouville
eigenvalue problem in the mathematics literature. We want to
approximate $E_1$ with relative error $\e$. 

Here, $\Delta$ is the $d$-dimensional Laplacian and $V\ge 0$
is a function of $d$
variables. The dimension is proportional to the number of particles, e.g.
$d=3p$. For many applications the number of particles $p$
and hence $d$ is huge. 
We consider algorithms that 
approximate $E_1$ using finitely many function evaluations of $V$.
Moreover, we assume that $V$ and its first
order partial derivatives $\partial V / \partial x_j$, $j=1,\dots,d$, are
continuous and uniformly bounded by~$1$. 

\section{Complexity of classical algorithms and discretization error}

Decades of calculating ground state eigenvalues of systems with
a large number of particles have suggested that such problems are hard.
We sketch a proof that the cost
of classical deterministic algorithms that approximate eigenvalues
in the worst case grows exponentially with the number of variables. 

Indeed, consider a potential function $V$ and let $\overline V$ be a 
perturbation of $V$.
Then the eigenvalue
$E_1(V)$ corresponding to $V$ and the eigenvalue $E_1(\overline V)$ corresponding
to $\overline V$ are related according to the formula
\begin{eqnarray*}
E_1(V) &=& E_1(\overline V) +
                  \int_{I_d} (V(x)-\overline V(x)) \Psi_{1}^2(x;\overline V)\,dx\\
       &+& O\big(\|V-\overline V\|_\infty^2\big),
\end{eqnarray*}
where $\Psi_1(\cdot ; \overline V)$ denotes the eigenfunction corresponding to
$E_1(\overline V)$. This implies that approximating $E_1$
is at least as hard as approximating a multivariate integral in the worst case.
As a result, any classical deterministic algorithm for the eigenvalue problem
with accuracy $\e$ must use a number of
function evaluations of $V$ that grows as
$\e^{-d}$; 
see \cite{Pap07} for details.
Determining whether the problem suffers the curse of dimensionality in the classical randomized setting
is an open problem. 
At the time of this writing we have an exponential gap between the known upper and lower randomized complexity bounds.

Finite differences are often used for approximating $E_1$. 
The discretization of the operator $-\tfrac 12 \Delta + V$ with
mesh size $h=(m+1)^{-1}$ yields an $m^d\times m^d$ matrix
$M_h:=M_h(V)= -\tfrac 12 \Delta_h + V_h$. Then one solves the
corresponding matrix eigenvalue problem $M_h z_{h,1} = E_{h,1}
z_{h,1}$.
Note that $\Delta_h$ denotes the discretization of the
Laplacian and $V_h$ is a diagonal matrix whose entries are
evaluations of the potential $V$ at the $m^d$ grid points. 
The reader may assume that $\Delta_h$ is obtained using a $2d+1$ 
stencil for the Laplacian; see e.g. \cite[p. 60]{LeVeque}.

For instance, if $d=2$ we have
\begin{equation*}
-\Delta_h = h^{-2} \left(
\begin{array}{rrrrr}
 T_h& -I \\
-I& T_h      & -I \\
  & \ddots & \ddots & \ddots \\
  &        & -I     & T_h  & -I \\
  &        &        & -I     & T_h
\end{array} \right),
\end{equation*}
is an $m^2\times m^2$ matrix,
where $I$ is the $m\times m$ identity matrix
while
\begin{equation*}
V_h = \left(
\begin{array}{rrrrr}
v_{11} \\
    & \ddots \\
    &        & v_{ij} \\
    &        &        &\ddots \\
    &        &        &       & v_{mm}
\end{array} \right) ,
\end{equation*}
where 
$v_{ij} = V(ih, jh)$, $i,j=1,\dots,m$, and $T_h$ is the
$m\times m$ matrix given by
\begin{equation*}
T_h= \left( \begin{array}{rrrrr}
 4  & -1 \\
 -1 &  4     & -1 \\
    & \ddots & \ddots & \ddots \\
    &        & -1     & 4      & -1 \\
    &        &        & -1     & 4
\end{array} \right) .
\end{equation*}

$M_h$ is
symmetric positive definite and sparse and has been extensively
studied in the literature \cite{Demmel,For04,LeVeque}. For $V$ that has
bounded first order partial derivatives, using
the results of \cite{Wein56,Wein58} we conclude 
\begin{equation}
\label{r1}
| E_1 - E_{h,1}| \le c_1 dh.
\end{equation}
If $\widehat E_{h,1}$ is such that $|E_{h,1}-\widehat E_{h,1}|
\le c_2dh$, we have relative error 
$$| 1 - \widehat E_{h,1}/E_1 | \le c^\prime h,$$
where $c^\prime$ is a constant.
The inequality follows by observing
that $2E_1$ is bounded from below by the smallest eigenvalue 
$4dh^{-2}\sin^2(\pi h/2)$ of the discretized Laplacian.

Such a discretization approach for a multiparticle system
is not new; see e.g. \cite[p. 621]{LeBris}. The problem
is that the size of the resulting matrix is exponential in $d$ and so is
the cost of classical algorithms approximating its ground state eigenvalue.

\section{Quantum algorithm}

We assume that $\e < 2/(d\pi^2)$ since otherwise we can approximate the smallest eigenvalue with relative 
error $\e$ with constant cost.
Indeed, for $V$ uniformly bounded by one, the smallest eigenvalue $E_1(V)$ 
satisfies $E_1(0) \le E_1(V) \le E_1(0) + 1$, where 
$E_1(0)=d\pi^2/2$ is the smallest eigenvalue of $-\tfrac 12 \Delta$.
Thus, 
$$\frac{| E_1(V) - E_1(0)|}{E_1(V)}\le \frac{1}{E_1(0)}= \frac {2}{d\pi^2}.$$
Therefore it suffices to deal only with the case $\e < 2/(d\pi^2)$.

First we discuss our algorithm in general terms and then we provide
a complete analysis. 
The key observation is that the discretization 
we outlined above and the estimation of the smallest eigenvalue
of the resulting matrix 
can be implemented on a quantum
computer with cost that does not grow exponentially with $d$.
This is accomplished by modifying  
quantum phase estimation, 
a well known quantum algorithm for approximating an
eigenvalue of a unitary matrix $W$, see e.g., \cite[p. 225]{NC00}.
First we provide a high level description of the algorithm and then 
give all its details and the resulting error and cost estimates.

\vskip 1pc 
\noindent{\bf Sketch of the algorithm:}
\begin{enumerate}
\item Consider the discretization $M_h=-\tfrac 12 \Delta_h + V_h$ of 
$-\tfrac 12 \Delta + V$
and let $h\le\e$ leading to the desired accuracy.
The matrix
$$W = e^{i M_h/(2 d)},$$
is unitary since $M_h$ is Hermitian.
\item For $W$ use
phase estimation to approximate the phase corresponding to
$e^{iE_{h,1}/(2d)}$ with the following 
modifications:
\begin{enumerate}
\item 
Use the approximate eigenvector
$$\ket{0}^{\otimes b}\ket{\psi_1}^{\otimes d}$$
as an initial state, where $\ket{\psi_1}^{\otimes d}$ is the ground state
eigenvector of $-\Delta_h$ and can be implemented efficiently; 
see the discussion following (\ref{eq:r5}) below for details.
\item Replace $W^{2^t}$, $t=0,\dots, b-1$, 
that are required in phase estimation, using approximations given by
high order splitting formulas that deal with the exponentials of
$-\tfrac 12 \Delta_h$ and $V_h$ separately and can be implemented efficiently;
see the discussion leading to (\ref{eq:r50}) below for details. 
\end{enumerate}
\end{enumerate}

The effect of the modifications is to somewhat decrease
the success probability while increasing 
the cost of phase estimation.
Nevertheless, the resulting success probability is at least $\tfrac 23$,
and the cost for implementing the initial state and the approximate
powers of $W$ does not suffer from the curse of dimensionality.
(The actual value of the success probability is not important 
since it exceeds $\tfrac 12$ and 
can be boosted to become arbitrarily close to one; 
see \cite[p. 153]{NC00} for details.) 

\begin{theorem}
Phase estimation with an approximate initial state and approximate
powers of $W$ with probability at least $\tfrac 23$ yields 
an estimate of $E_1$ with relative error $\e$ and 
total cost 
$$C d\, \e^{-(3+\delta)},$$ 
for any  $\delta > 0$, 
using $C^\prime d\log \e^{-1}$ qubits, where $C$ and $C^\prime$ are constants.
The pseudocode for the algorithm is given in listing Algorithm~\ref{alg:grste}.
\end{theorem}

\begin{algorithm}
\caption{GroundStateEnergy($\e$, $d$, $V$)}
\label{alg:grste}
\begin{algorithmic}[1]
\REQUIRE $d$ to be a positive integer.
\REQUIRE $\e \in (0,2/\pi^2 d)$. Note that for relative error $\e \ge 2/\pi^2 d$
the problem can be solved with constant cost.
\REQUIRE $V:[0,1]^d\rightarrow [0,1]$ to be provided by an oracle (black box).
\STATE $b \leftarrow \lceil -\log_2 \e \rceil$
\STATE $m \leftarrow 2^{b} - 1$
\STATE $h \leftarrow (m+1)^{-1}$
\STATE Initial state : $\ket{0}^{\otimes b}\ket{\psi_1}^{\otimes d}$
\COMMENT{The right register holds the eigenvector of the discretized Laplacian, with mesh size $h$. The corrdinates of
$\ket{\psi_1}$ are given in equation (\ref{eq:r5-1})}
\STATE ApproxW($b$, $h$, $m$, $d$, $V_h$, $\widetilde W$)
\COMMENT {This subroutine call returns $\widetilde W$ which is a list of the approximations of the exponentials 
$W^{2^t}= e^{(-\tfrac12 \Delta_h + V_h)2^t/(2d)}$, $t=0,\dots b-1$. These approximations are denoted by  
$\widetilde{W}_t$, $t=0,\dots, b-1$; see Algorithm~\ref{alg:ApproxW} for details.}
\STATE Apply Hadamard to the left $b$-qubit register:
\[\left(H^{\otimes d}\otimes I\right)\ket{0}^{\otimes d}\ket{\psi_1}^{\otimes d}\rightarrow \frac 1{2^{b/2}} \sum_{i_0,i_1,\dots,i_{b-1}=0}^1 \ket{i_{b-1}i_{b-2}\cdots i_1i_0} \ket{\psi_1}^{\otimes d}\]
\STATE Apply $\widetilde W_0$\dots $\widetilde W_{b-1}$, controlled by the left $b$ qubits, 
to the state above: 
$$\rightarrow
\frac 1{2^{b/2}} \sum_{i_0,i_1,\dots,i_{b-1}=0}^1
\ket{i_{b-1}j_{b-2}\cdots i_1i_0}
\widetilde W_{b-1}^{i_{b-1}}\cdots \widetilde W_1^{i_1} \widetilde W_0^{i_0} \ket{\psi_1}^{\otimes d}$$
\STATE Apply the inverse Fourier transform $FT^\dag$ to the register holding the leftmost $b$ qubits: \[\rightarrow
(FT^\dag \otimes I) \left( \frac 1{2^{b/2}} \sum_{i_0,i_1,\dots,i_{b-1}=0}^1
\ket{i_{b-1}i_{b-2}\cdots i_1i_0}
\widetilde W_{b-1}^{i_{b-1}}\cdots \widetilde W_1^{i_1} \widetilde W_0^{i_0} \ket{\psi_1}^{\otimes d} \right)\]
\STATE Measure the first $b$ qubits in the computational basis: outcome ($j_{b-1},\dots,j_1,j_0)$
\STATE $j \leftarrow \sum_{k=0}^{b-1} j_k 2^k$
\STATE $\widehat E_{h,1} \leftarrow 4\pi d j /2^b$
\RETURN $\widehat E_{h,1}$
\end{algorithmic}
\end{algorithm}

\begin{algorithm}
\caption{ApproxW($d$, $b$, $h$, $m$, $V$, $\widetilde{W}$)}
\label{alg:ApproxW}
\begin{algorithmic}[1]
\REQUIRE $d$ is a positive integer. 
\REQUIRE $b$ is a positive integer defined in Algorithm~\ref{alg:grste}.
\REQUIRE $h$ is a positive real number defined in Algorithm~\ref{alg:grste}.
\REQUIRE $m$ is a positive integer defined in Algorithm~\ref{alg:grste}.
\REQUIRE $V:[0,1]^d\rightarrow [0,1]$ to be provided by an oracle (black box).
\REQUIRE $\widetilde{W}$ to be a list where this subroutine will hold the approximations 
$\widetilde{W}_t$, $t=0,\dots,b-1$, that it computes. This list is returned to the calling program.
\STATE Let $V_h$ be the $m^d\times m^d$ diagonal matrix obtained by discretizing the function $V$ 
on a grid with mesh size $h = (m+1)^{-1}$. 
\STATE $\textrm{Norm}_1 \leftarrow h^{-2} \sin^2 ( \frac{\pi m}{2(m+1)} )$
\COMMENT {The norm of $H_1= - \Delta_h/(4d)$.}
\STATE $\textrm{Norm}_2 \leftarrow 1/(2d)$
\COMMENT {Upper bound of the norm of $H_2 = V_h/(2d)$.}
\STATE $k \leftarrow \lfloor \sqrt{\frac{1}{2}\log_{25/3}\frac{80e2^b}{d}} + \frac{1}{2}\rfloor$ 
\COMMENT {Note $k\ge 1$ by definition of $b$.}
\STATE $c_k \leftarrow \frac{8}{3}k\left(\frac{5}{3}\right)^{k-1}$
\COMMENT {See also \cite[Eq. 7]{PZ10}.}
\STATE $\mathcal{H}_1 = \frac{-\Delta_h}{4d\cdot \textrm{Norm}_1}$
\STATE $\mathcal{H}_2 = \frac{V_h}{2d\cdot \textrm{Norm}_1}$
\FOR {$t=0$ to $b-1$}
\STATE $\e_t \leftarrow 2^{t+1-b}/40$
\STATE $M \leftarrow \left(\frac{8e2^t\textrm{Norm}_2}{\e_t}\right)^{1/(2k)}\frac{2ec_k}{2k+1}$
\STATE ${\rm NumberOfIntervals} \leftarrow \lceil M \; \textrm{Norm}_1 \;2^t \rceil$
\COMMENT{This is the number of subintervals the intervals we divide the interval $[0, \textrm{Norm}_1 \; 2^t]$.}
\STATE ${\rm IntervalSize} \leftarrow \frac {\textrm{Norm}_1 \; 2^t}{\rm NumberOfIntervals}$
\COMMENT {Each subinterval has size at most $1/M$.}
\STATE $p_k \leftarrow (4-4^{1/2k-1})^{-1}$
\STATE $S_{2}({\rm IntervalSize}) \leftarrow e^{-i\mathcal{H}_1 {\rm IntervalSize} /2}
e^{-i\mathcal{H}_2 {\rm IntervalSize}} e^{-i\mathcal{H}_1 {\rm IntervalSize} /2}$ 
\FOR {$j=2$ to $k$}
\STATE Let
\begin{eqnarray*}
S_{2j}({\rm IntervalSize}) &\leftarrow&
\left[ S_{2j-2} (p_k {\rm IntervalSize})\right]^2 
\left[ S_{2j-2} ((1-4p_k ) {\rm IntervalSize})\right] \\
&\times& \left[ S_{2j-2} (p_k {\rm IntervalSize})\right]^2 
\end{eqnarray*}
\ENDFOR
\STATE $\widetilde{W}_t \leftarrow [S_{2k} ({\rm IntervalSize})]^{\rm NumberOfIntervals}$
\ENDFOR
\RETURN $\widetilde{W}_t$, $t=0,\dots,b-1$, as the list $\widetilde{W}$
\end{algorithmic}
\end{algorithm}

Next we discuss the details of our algorithm and this will lead us
to the proof of the theorem.
Let $h=(m+1)^{-1}$, where $m= 2^{\lceil -\log_2 \e \rceil} - 1$. Clearly, $h\le \e  < 2/(d\pi^2) < 1/4$
due to our assumption at the beginning of this section.
This leads to the desired accuracy while
ensuring the discretization is not trivial. 
The eigenvalue of $W$ that
corresponds to $E_{h,1}$ is $e^{i E_{h,1}/ (2 d)} =
e^{2\pi i \varphi_1}$, where
$$\varphi_1 = E_{h,1} / (4\pi d)$$
is the phase and belongs to the interval
$[0,1)$ since $E_{h,1} \le 2 d h^{-2} \sin^2 (\pi h/2) + 1 \le
d\pi^2/2 + 1$. 

Quantum phase estimation approximates the phase
$\varphi_1$ with $b$-bit accuracy, where $b = \lceil -\log_2 \e \rceil$. 
The output of the algorithm is an index $j\in [0, 2^b-1]$ such that
$
| \varphi_1 -  j\, 2^{-b} | \le  2^{-b}.
$
Hence,
\begin{equation}
\label{r2} | E_{h,1} -  4\pi dj\, 2^{-b} | \le c_2 d\e.
\end{equation}
Combining (\ref{r1}) and (\ref{r2}) we conclude
\begin{equation}
\label{r3} | E_{1} - 4\pi dj\, 2^{-b} | \le c_1 d \e
+ c_2 d \e = c d\e.
\end{equation}
Hence the algorithm approximates the ground state eigenvalue $E_1$ by 
$$\widehat E_{h,1}:=  4\pi dj\, 2^{-b}.$$

The estimate $\widehat E_{h,1}$ holds with probability at least $\tfrac 8{\pi^2}$ (see, e.g.,
\cite{AmAm}) assuming:
\begin{itemize} 
\item The initial state of the algorithm is
$\ket{0}^{\otimes b}\ket{z_{h,1}}$, where $\ket{z_{h,1}}$ is the
eigenvector of $M_h$ that corresponds to $E_{h,1}$. 
\item We are given the matrix exponentials $W^{2^t}$,
$t=0,\dots,b-1$.
\end{itemize}

In our case, however, we do not know $\ket{z_{h,1}}$ and
we use an approximation. 
Similarly, we use approximations of the
$W^{2^t}$, $t=0,\dots, b-1$, to simulate the evolution of the
quantum system that evolves with Hamiltonian $H= M_h / (2d)$. We
will compute the cost to implement these approximations so that
(\ref{r3}) holds. All these approximations affect the estimate $\tfrac
8{\pi^2}$ of the success probability of phase estimation, but only
by a small amount. 

The initial state of our algorithm is
\begin{equation}\label{eq:r5}
\ket{0}^{\otimes b}\ket{\psi_1}^{\otimes d},
\end{equation}
where $\ket{\psi_1}^{\otimes d}$
is  the ground state eigenvector of the discretized Laplacian.
We know \cite{Demmel} that the coordinates of $\ket{\psi_1}$ are
\begin{equation}\label{eq:r5-1}
\psi_{1j} = \sqrt{2h} \sin(j\pi h),\quad j=1,\dots,m,
\end{equation}
and $\ket{\psi_1}^{\otimes d}$ has unit length. 
Since $h$ is proportional to $\e$, the matrix $M_h$ has size
$m^d\times m^d$, with $m=\Theta(\e^{-1})$. 
Therefore,
$\ket{\psi_1}^{\otimes d}\in\complex^{m^d}$ 
and can be  
represented using $\log_2 m^d = O(d
\log_2 \e^{-1})$ qubits and can be implemented with
$d\cdot O(\log^2 \e^{-1})$ quantum operations using the 
Fourier transform; see e.g., \cite{KR01,Wick94}. 
We point out 
that here and elsewhere the implied constants in the big-$O$ 
and $\Theta$ notation 
are independent of $d$ and $\e$.
(From a practical standpoint, it is possible to further reduce the cost
of the initial state using the algorithm in \cite{JP03} but we do not pursue
this alternative since the analysis of the algorithm becomes more involved.)

Expanding $\ket{\psi_1}^{\otimes d}$ using the eigenvectors of $M_h$ we have
$$\ket{\psi_1}^{\otimes d} = \sum_{k=1}^{m^d}d_k\ket{z_{h,k}}.$$
The approximate initial state reduces
the success probability of phase estimation
by a factor equal to the square of the magnitude of the projection of
$\ket{\psi_1}^{\otimes d}$ onto $\ket{z_{h,1}}$,
to become
$\frac{8}{\pi^2}|d_1|^2$; see, e.g., \cite{AL99,JP03}.

We will see that $|d_1|^2 > \pi^2/10$. Indeed,
we estimate $|d_1|$ using the approach in \cite[p. 172]{Wilkinson} which
is based on the separation of the eigenvalues of $M_h$. In particular, we have
$$1\ge (E_{h,2} - E_{h,1})^2(1 - |d_1|^2),$$
where $E_{h,1}$ and $E_{h,2}$ are the smallest and second smallest 
eigenvalues of $M_h$.
We estimate $E_{h,2} - E_{h,1}$ from below using the two smallest
eigenvalues of $-\Delta_h$ to obtain
$E_{h,2} - E_{h,1} \ge 2h^{-2}(\sin^2(\pi h) - \sin^2(\pi h/2)) -1.$

This yields that the success probability of phase estimation 
with the approximate ground state eigenvector is at least
\begin{equation}
\label{r4}
\frac{8}{\pi^2}\left( 1-
\frac{1}{(2h^{-2}(\sin^2(\pi h) - \sin^2(\pi h/2)) -1)^2} \right) > \frac 45,
\end{equation}
$h\le 1/4$. (The overall success probability of the algorithm is affected
by an additional factor and once we address that we will provide
a final estimate.)

Now let us turn to the approximation of the matrix exponentials. We
simulate the evolution of a quantum system with Hamiltonian $H =
M_h/(2 d)$ for time $2^t$, $t=0,1,\dots, b-1$. Let $H = H_1 + H_2$
where $H_1 = -\Delta_h / (4 d)$ and $H_2 = V_h/ (2 d)$. Recall that
$h$ is the largest mesh size satisfying $h\le\min(\e,1/4)$.
The eigenvalues and eigenvectors of the
discretized Laplacian are known and the evolution of a system with
Hamiltonian $H_1$ can be implemented with $d \cdot O(\log^2 \e^{-1})$
quantum operations using the Fourier transform in each dimension; see e.g.,
\cite[p. 209]{NC00}. The evolution of a system with Hamiltonian $H_2$ can be
implemented using two quantum queries and phase kickback.
The queries are similar to those in Grover's algorithm \cite{NC00} and return
function evaluations of $V$ truncated to
$O(\log\e^{-1})$ bits. 

In particular, we use a splitting formula $S_{2k}$ of order $2k+1$, $k\ge 1$, to
approximate $W^{2^t}=e^{i (H_1+H_2)2^t}$ by a product of the form
\begin{equation}\label{eq:r50}
\prod_{\ell=1}^{N_t} e^{i A_\ell z_\ell},
\end{equation}
where $A_\ell \in\{H_1,H_2\}$ and suitable $z_\ell$ that depend on $t$ and $k$.

The splitting formula $S_{2k}$ is due to Suzuki \cite{Su90,Su91}. It is used to approximate 
$e^{i(B+C)\Delta t}$, where $B$ and $C$ are Hermitian matrices. This formula is defined recursively by
\begin{eqnarray*}
S_2(B,C,\Delta t) &=& e^{iB\Delta t/2} e^{iC\Delta t} e^{iB\Delta t/2} \\
S_{2k}(B,C,\Delta t) &=& [S_{2k-2}(B,C, p_k\Delta t) ]^2 S_{2k-2}(B,C, (1-4p_k) \Delta t) \\
&& \hspace{10pc} \times [S_{2k-2}(B,C, p_k\Delta t) ]^2,
\end{eqnarray*}
where $p_k= (4 - 4^{1/(2k-1)})^{-1}$, $k=2,3,\dots$. 

Unfolding the recurrence above and combining it with \cite[Th. 1]{PZ10} we obtain that the approximation of $W^{2^t}$ has the form
\begin{equation}
\label{eq:unfold}
\widetilde W^{2^t} = e^{iH_1 a_{t,0}} e^{iH_2 b_{t,1}} e^{i H_1 a_{t,1}} \cdots e^{iH_2 b_{t,L_t}} e^{i H_1 a_{t,L_t}},
\end{equation}
where
$s_{t,0},\dots, s_{t,L_t}$ and $z_{t,1},\dots,z_{t,L_t}$ and $L_t$ are parameters, $t=0,\dots, b-1$. 
The number of exponentials involving $H_1$ and $H_2$ in the expression above is $N_t=2L_t+1$.
The precise definition of the $\widetilde W^{2^t}$, $t=0,\dots,b-1$, 
is presented in pseudocode listing Algorithm~\ref{alg:ApproxW}. 

Let $\| \cdot\|$ be the matrix norm induced by the Euclidean vector norm.
From \cite[Thm. 1 \& Cor. 1]{PZ10} the number $N_t$
of exponentials needed to approximate $W^{2^t}$
by a splitting formula of order $2k+1$ with
error $\e_t$, $t=0,\dots,b-1$, is
\begin{equation*}
N_t \leq  16e \|H_1\| 2^t\, \left(\frac {25}3\right)^{k-1}
\left(\frac{8e\,2^t\|H_2\|}{\e_t}\right)^{1/(2k)}, 
\end{equation*}
for any $k \geq 1$.
The total number of exponentials required for
the approximation of all the $W^{2^t}$ is bounded from above as follows
\begin{eqnarray}
N&=&\sum_{t=0}^{b-1} N_t 
\le 16e \| H_1\| \left(\frac {25}3\right)^{k-1}
 \left(8e\|H_2\|\right)^{1/(2k)} \nonumber \\
&& \quad\quad\quad\quad\times\;   \sum_{t=0}^{b-1}2^t
\left( \frac {2^t}{\e_t}\right)^{1/(2k)} \label{eq:numexp} \\
&\le& 16e \| H_1\| 2^b \left(\frac {25}3\right)^{k-1}
\left( 160e\,2^b\|H_2\| \right)^{1/(2k)},
 \nonumber
\end{eqnarray}
where we obtained the last inequality by setting
$\e_t = \tfrac {2^{t+1 - b}}{40}$,
$t=0,\dots,b-1$. It is easy to check that
$\sum_{t=0}^{b-1}\e_t \le \tfrac 1{20}$.
Thus the success probability
of phase estimation can be reduced by twice this amount \cite[p. 195]{NC00}. 
Using (\ref{r4}) we
conclude our algorithm succeeds with probability at least
$$ \frac 45 - \frac 1{10} > \frac 23.$$

The largest eigenvalue of $-\Delta_h$ is $4dh^{-2} \sin^2(\pi m h/2)$. Since $H_1 = - \Delta_h/ (4d)$ we have
$\norm{H_1} \leq \frac{4dh^{-2}}{4d} \leq \e^{-2}$. Since $V$ is uniformly bounded by one and $H_2 = V_h/(2d)$ we
have 
$\norm{H_2} \leq 1/(2 d)$. Hence, the algorithm uses a number of
exponentials of $H_1$ and $H_2$ that satisfies
\begin{equation*}
N \le 
16e
\left(\frac{80e}{d}\right)^{1/(2k)} 
\left(\frac {25}3\right)^{k-1}  \e^{-2} \; 2^{b(1+1/(2k))}.
\end{equation*}
Since we have chosen $b = \lceil -\log_2 \e \rceil$ we
obtain
$$
N \le \widetilde C\;
\left(\frac{80e}{d}\right)^{1/(2k)}
\left(\frac {25}3\right)^{k-1}\; \e^{-(3+\frac{1}{2k})},
$$
for any $k>0$, where $\widetilde C$ is a constant.

The {\it optimal} $k^*$, i.e., the one minimizing the upper bound for $N$
in (\ref{eq:numexp}), is obtained in \cite[Sec. 5]{PZ10} and is given by
\[
k^* = \left\lfloor \sqrt{\frac{1}{2}  \log_{25/3} \frac{80e\;2^b}{d}} + \frac 12 \right\rfloor = 
O\left( \sqrt{\ln \frac 1{d\e}}\right)
\quad {\rm as\ }d\e \to 0,
\]
by definition of $b$.
The number of exponentials corresponding to $k^*$ satisfies
\begin{equation}\label{eq:numqueries}
N^{*} = O\left(\e^{-3}e^{\sqrt{\ln \frac{1}{d\e}}}\right) 
\quad {\rm as\ } d\e\to 0.
\end{equation}

We remark that of the $N^*$ matrix exponentials roughly half involve $H_1$
and the remaining involve $H_2$; see (\ref{eq:unfold}).
Since each exponential involving $H_2$
requires two queries the total number of queries is also $N^*$.

Hence,
the number of quantum operations, excluding queries, to implement the
initial state, the  matrix exponentials involving $H_1$ and the inverse
Fourier transform yielding the final state of phase estimation is
\begin{equation}\label{eq:numops}
N^* \cdot O(d\log^2\e^{-1}).
\end{equation}

Equations (\ref{eq:numexp}), (\ref{eq:numqueries}) and (\ref{eq:numops}) yield 
that the total cost of the algorithm, including the number of
queries and the number of all other quantum operations, is
$$Cd\e^{-(3+\delta)},$$
where $\delta >0$ is arbitrarily small and $C$ is a constant.

Summarizing our results 
we see that the dependence on $d$ of the number of qubits and the cost
is linear.
As far as the number of qubits is concerned this is not really surprising.
The algorithm uses phase estimation to approximate 
an eigenvalue of a matrix whose size is
proportional to $\e^{-d}\times \e^{-d}$. The corresponding eigenvector
has a number of coordinates proportional to $\e^{-d}$ and, therefore,
is represented using a number of qubits proportional to $d\log_2 \e^{-1}$.

We now turn to the cost. 
The depth of the quantum circuit realizing the algorithm grows as
$N^*$ which is given in (\ref{eq:numqueries}).
Clearly, $\e^{-3}e^{\sqrt{\ln \frac{1}{d\e}}} \le
\e^{-3}e^{\sqrt{\ln \frac{1}{\e}}}$,
for any $d$. Thus $N^*$ is bounded from above by a quantity independent of $d$.
Recall that $N^*$ is
the total number of matrix exponentials the algorithm uses.
Roughly half of these exponentials 
involve the discretized Laplacian $\Delta_h$ and the rest
involve the discretized potential $V_h$. 

Each of the matrix exponentials involving the $d$ dimensional $\Delta_h$ 
is implemented efficiently with cost proportional to
$d\log^2\e^{-1}$ using the quantum Fourier transform. 
Hence the cost of all matrix exponentials involving $\Delta_h$
depends linearly on $d$.

We consider the cost of the matrix exponentials involving $V_h$.
Each exponential can be implemented with two quantum queries. 
We assume the cost of each query is constant.
Hence the cost of all matrix exponentials involving $V_h$
is $2N^*$ times the cost of a quantum query.

Thus the sum of the cost of all matrix exponentials and, therefore,
the cost of the algorithm depends linearly on $d$.

This cost analysis has the advantage that it reveals the computational
effort spent on solving the ground state eigenvalue problem unobscured
by the actual cost of evaluating $V$ (i.e., the the cost of a quantum query).
It is not limited in any way, since
for any particular choice of
$V$ when the actual cost of a query is known, it 
suffices to multiply it by the number of queries and
add the product to (\ref{eq:numops}) to
obtain an aggregate cost estimate.

For multiparticle systems studied in physics and chemistry the number 
of dimensions $d$ is directly proportional to the number of particles $p$.
For instance, $p$ particles in three dimensions yield $d=3p$. 
Thus the dependence on $p$ of the number of qubits and the cost 
of the algorithm is linear.

Finally, our analysis assumes
a perfect physical realization of a quantum computer.
However, for the implementation of the algorithm, one needs to address 
decoherence and other sources of error
for a specific underlying architecture.
This may significantly increase the required computational resources. 
Such a study exists for phase estimation and the Abrams and 
Lloyd algorithm  \cite{AL99} 
applied to the ground state eigenvalue of the transverse Ising 
model~\cite{8}; see also the references therein and \cite{9}.
This study is broad enough to cover Shor's algorithm 
and conveys the general idea in our case as well.
It concludes that for the current state of the art in quantum logic array 
architectures the existing fault tolerance and error correction techniques
impose significant resource requirements in the implementation of these
algorithms.

{\it Acknowledgements.}
Joseph Traub would like to thank the Santa Fe Institute for its endlessly
stimulating environment.
We thank Rolando Somma, LANL, for his comments.
This work has been supported in part by the National Science Foundation.

\bibliographystyle{plain}
\bibliography{grstateRevised}

\begin{thebibliography}{10}

\bibitem{AL99}
D.~S. Abrams and S.~Lloyd.
\newblock Quantum algorithm providing exponential speed increase for finding
  eigenvalues and eigenvectors.
\newblock {\em Phys.\ Rev.\ Lett.}, 83:5162--5165, 1999.

\bibitem{var}
C.~H. Bennet, E.~Bernstein, G.~Brassard, and U.~Vazirani.
\newblock Strengths and weaknesses of quantum computing.
\newblock {\em SIAM J. Computing}, 26:1510--1523, 1997.

\bibitem{AmAm}
G.~Brassard, P.~Hoyer, M.~Mosca, and A.~Tapp.
\newblock {\em Quantum Amplitude Amplification and estimation}, volume 305,
  page~53.
\newblock In Contemporary Mathematics, Quantum Computation and Information,
  Samuel J. Lomonaco Jr. and Howard E. Brandt, Editors, AMS, Providence, RI,
  2002.
\newblock http://arXiv.org/abs//quant-ph/0005055.

\bibitem{5}
S.~Bravyi, D.~DiVincenzo, and D.~Loss.
\newblock Polynomial-time algorithm for simulation of weakly interacting
  quantum spin systems.
\newblock {\em Communications in Mathematical Physics}, 284:481--507, 2008.

\bibitem{LeBris}
P.~G. Ciarlet and C.~Le Bris.
\newblock {\em Handbook of Numerical Analysis, Special Volume Computational
  Chemistry}, volume~X.
\newblock North Holland, Amsterdam, 2003.

\bibitem{8}
C.~R. Clark, T.~S. Metodi, S.~D. Gasster, and K.~R. Brown.
\newblock Resource requirements for fault-tolerant quantum simulation: The
  ground state of the transverse ising model.
\newblock {\em Phys. Rev. A}, 79:062314, 2009.

\bibitem{Demmel}
J.~W. Demmel.
\newblock {\em Applied Numerical Linear Algebra}.
\newblock SIAM, Philadelphia, PA, 1997.

\bibitem{JDu10}
J.~Du, N.~Xu, X.~Peng, P.~Wang, S.~Wu, and D.~Lu.
\newblock {NMR} implementation of a molecular hydrogen quantum simulation with
  adiabatic state preparation.
\newblock {\em Phys. Rev. Lett.}, 104:030502, 2010.

\bibitem{For04}
G.~E. Forsythe and W.~R. Wasow.
\newblock {\em Finite-Difference Methods for Partial Differential Equations}.
\newblock Dover, New York, 2004.

\bibitem{9}
H.~H\"affner, C.~F. Roos, and R.~Blatt.
\newblock Quantum computing with trapped ions.
\newblock {\em Phys. Reports}, 469:155, 2008.

\bibitem{1}
A.~Hams and H.~DeRaedt.
\newblock Fast algorithm for finding the eigenvalue distribution of very large
  matrices.
\newblock {\em Phys. Rev. E}, 62(3):4365--4377, 2000.

\bibitem{JP03}
P.~Jaksch and A.~Papageorgiou.
\newblock Eigenvector approximation leading to exponential speedup of quantum
  eigenvalue estimation.
\newblock {\em Phys. Rev. Lett.}, 91:257902, 2003.

\bibitem{Kassal08}
I.~Kassal, S.~P. Jordan, P.~J. Love, M.~Mohseni, and A.~Aspuru-Guzik.
\newblock Polynomial-time quantum algorithm for the simulation of chemical
  dynamics.
\newblock {\em PNAS}, 105:18681--18686, 2008.

\bibitem{Kassal10}
I.~Kassal, J.~D. Witfield, A.~Perdomo-Ortiz, Man-Hong Yung, and
  A.~Aspuru-Guzik.
\newblock Quantum information and computation for chemistry.
\newblock {\em An. Rev. Phys. Chem.}, 62:185--207, 2011.
\newblock http://arxiv.org/abs/1007.2648.

\bibitem{kit}
J.~Kempe, A.~Kitaev, and O.~Regev.
\newblock The complexity of the local hamiltonian problem.
\newblock {\em SIAM J. Computing}, 35:1070--1097, 2006.

\bibitem{KR01}
A.~Klappenecker and M.~R\"otteler.
\newblock Discrete cosine transforms on quantum computers, 2001.
\newblock http://arXiv.org/quant-ph/0111038.

\bibitem{nature}
B.~P. Lanyon, J.~D. Whitfield, G.~G. Gillett, M.~E. Goggin, M.~P. Almeida,
  I.~Kassal, J.~D. Biamonte, M.~Mohseni, B.~J. Powell, M.~Barbieri,
  A.~Aspuru-Guzik, and A.~G. White.
\newblock Towards quantum chemistry on a quantum computer.
\newblock {\em Nature Chemistry}, 2:106--111, 2010.

\bibitem{LeVeque}
R.~J. LeVeque.
\newblock {\em Finite Difference Methods for Ordinary and Partial Differential
  Equations}.
\newblock SIAM, Philadelphia,PA, 2007.

\bibitem{Lub}
C.~Lubich.
\newblock {\em From Quantum to Classical Molecular Dynamics: Reduced Models and
  Numerical Analysis}.
\newblock European Mathematical Society, Z\"urich, 2008.

\bibitem{NC00}
M.~A. Nielsen and I.~L. Chuang.
\newblock {\em Quantum Computation and Quantum Information}.
\newblock Cambridge University Press, Cambridge, UK, 2000.

\bibitem{No01}
E.~Novak.
\newblock Quantum complexity of integration.
\newblock {\em J. Complexity}, 19:19--42, 2001.

\bibitem{NSW04}
E.~Novak, I.~H. Sloan, and H.~Wo\'zniakowski.
\newblock Tractability of approximation for weighted {K}orobov spaces on
  classical and quantum computers.
\newblock {\em Journal of Foundations of Computational Mathematics},
  4:121--156, 2004.

\bibitem{6}
S.~Oh.
\newblock Quantum computational method of finding the ground-state energy and
  expectation values.
\newblock {\em Phys. Rev. A}, 77:012326, 2008.

\bibitem{Pap07}
A.~Papageorgiou.
\newblock On the complexity of the multivariate {S}turm-{L}iouville eigenvalue
  problem.
\newblock {\em J. Complexity}, 23:802--827, 2007.

\bibitem{PZ10}
A.~Papageorgiou and C.~Zhang.
\newblock On the efficiency of quantum algorithms for {H}amiltonian simulation.
\newblock {\em Quantum Information Processing}, 11(2):541--561, 2012.
\newblock Online First, DOI: http://dx.doi.org/10.1007/s11128-011-0263-9.

\bibitem{Su90}
M.~Suzuki.
\newblock Fractal decomposition of exponential operators with applications to
  many body theories and monte carlo simulations.
\newblock {\em Phys. Lett. A}, 146:319--323, 1990.

\bibitem{Su91}
M.~Suzuki.
\newblock General theory of fractal path integrals with application to
  many-body theories and statistical physics.
\newblock {\em J. Math. Phys.}, 32:400--407, 1991.

\bibitem{4}
T.~Szkopek, V.~Roychowdhury, E.~Yablonovitch, and D.~S. Abrams.
\newblock Eigenvalue estimation of differential operators with a quantum
  algorithm.
\newblock {\em Phys. Rev. A}, 72:062318, 2005.

\bibitem{TW02}
J.~F. Traub and W.~Wo\'zniakowski.
\newblock Path integration on a quantum computer.
\newblock {\em Quantum Information Processing}, 1:365--388, 2002.

\bibitem{3}
P.~Varga and B.~Apagyi.
\newblock Phase estimation procedure to solve quantum-mechanical eigenvalue
  problems.
\newblock {\em Phys. Rev. A}, 78:022337, 2008.

\bibitem{7}
H.~Wang, S.~Ashhab, and F.~Nori.
\newblock Efficient quantum algorithm for preparing molecular-system-like
  states on a quantum computer.
\newblock {\em Phys. Rev. A}, 79:042335, 2009.

\bibitem{2}
H.~Wang, S.~Kais, A.~Aspuru-Guzik, and M.~R. Hoffmann.
\newblock Quantum algorithm for obtaining the energy spectrum of molecular
  systems.
\newblock {\em Phys. Chem. Chem. Phys.}, 10:5388--5393, 2008.

\bibitem{Wein56}
H.~F. Weinberger.
\newblock Upper and lower bounds for eigenvalues by finite difference methods.
\newblock {\em Comm. Pure Appl. Math}, 9:613--623, 1956.

\bibitem{Wein58}
H.~F. Weinberger.
\newblock Lower bounds for higher eigenvalues by finite difference methods.
\newblock {\em Pacific J. Math.}, 8:339--368, 1958.

\bibitem{Wick94}
M.~V. Wickerhauser.
\newblock {\em Adapted wavelet analysis from theory to software}.
\newblock A.K. Peters, Wellesley, MA, 1994.

\bibitem{Wilkinson}
J.~H. Wilkinson.
\newblock {\em The Algebraic Eigenvalue Problem}.
\newblock Oxford University Press, Oxford, UK, 1965.

\end{thebibliography}

\vskip 2pc
\noindent{\bf Authors' Addresses:}

\vskip 2pc
\noindent
A. Papageorgiou, \\
Department of Computer Science, \\
Columbia University,\\
New York, NY 10027, USA, email: ap@cs.columbia.edu\\
\smallskip

\noindent
I. Petras, \\
Department of Computer Science, \\
Columbia University,\\
New York, NY 10027, USA, email: ipetras@cs.columbia.edu\\
\smallskip

\noindent
J. F. Traub \\
Department of Computer Science, \\
Columbia University,\\
New York, NY 10027, USA, email: traub@cs.columbia.edu\\
\smallskip

\noindent
C. Zhang \\
Department of Computer Science, \\
Columbia University,\\
New York, NY 10027, USA, email: czhang@cs.columbia.edu

\end{document}